\documentclass[aps,pre,twocolumn,showpacs,superscriptaddress,groupedaddress]{revtex4}  
\pdfoutput=1
\usepackage{graphicx}  
\usepackage{dcolumn}   
\usepackage{bm}        
\usepackage{amssymb}   
\usepackage{amsmath}
\usepackage{color}
\usepackage{subfig}
\begin{document}


\title{A homotopic mapping between current and conductance-based synaptic mechanisms in a mesoscopic neural model }

%
\author{A.D.H.~Peterson} \affiliation{Department of Medicine, St.Vincent's Hospital, University of Melbourne, Australia}
\author{H.~Meffin} \affiliation{National Vision Research Institute, Australian College of Optometry} 
\author{M.J.~Cook}  \affiliation{Department of Medicine, St.Vincent's Hospital, University of Melbourne, Australia} 
\author{D.B.~Grayden} \affiliation{NeuroEngineering Lab, Department of Biomedical Engineering, University of Melbourne, Australia}  
\author{I.M.Y.~Mareels} \affiliation{IBM Research, Australia}
\author{A.N.~Burkitt} \affiliation{NeuroEngineering Lab, Department of Biomedical Engineering, University of Melbourne, Australia}
\vskip 0.25cm

\begin{abstract}
\noindent
Changes in brain states, as found in many neurological diseases such as epilepsy, are often described as bifurcations in mesoscopic neural models.  Nearly all of these models rely on a mathematically convenient, but biophysically inaccurate, description of the synaptic input to neurons called current-based synapses. We develop a novel analytical framework to analyze the effects of a more biophysically realistic description, known as conductance-based synapses. These are implemented in a mesoscopic neural model and compared to the standard approximation via a single parameter homotopic mapping. A bifurcation analysis using the homotopy parameter demonstrates that if a more realistic synaptic coupling mechanism is used in this class of models, then a bifurcation or transition to an abnormal brain state does not occur in the same parameter space.  We show that the more realistic coupling has additional mathematical parameters that require a fundamentally different biophysical mechanism to undergo a state transition. These results demonstrate the importance of incorporating more realistic synapses in mesoscopic neural models and challenge the accuracy of previous models, especially those describing brain state transitions such as epilepsy.   
\end{abstract}

\maketitle

\noindent
\textit{Introduction:}  
Modelling biological phenomena often involves mathematical descriptions of interacting nonlinear systems whose complex dynamics are shaped by feedback and noise processes.  Unlike in many physical systems such as condensed matter, there has been little progress in linking the dynamics of different spatiotemporal scales in neuroscience, which is still an open problem. In this paper, we examine the dynamic effects of a sub-cellular structure -- the synapse -- on the mesoscopic network behavior of neurons. A synapse connects or couples a \textit{pre-synaptic} neuron to a \textit{post-synaptic} neuron. When a signal arrives from the pre-synaptic neuron, a current is generated in the post-synaptic neuron. Biophysically, there is a flow of ions modulated by the membrane potential in accordance with Ohm's law.  This is known as a conductance-based synapse. Nearly all mesoscopic neural models mathematically approximate this as an injected current that is independent of the membrane potential.  This simplification is known as a current-based synapse and is considerably less biophysically accurate.  
\\ 
Modelling mesoscopic brain dynamics ($10^5$-$10^6$ neurons) typically employs the use of neural mass or neural field models \cite{Deco2008, Bressloff2012} inspired by mean-field theory.  These are low-dimensional, phenomenological, and describe the average behaviour of populations of neurons by their average firing rates $\phi(t)$ and  membrane potentials $V(t)$.  These models can reproduce normal and epileptic electroencephalography (EEG) signals \cite{Breakspear2006c}.  While there exist models with conductance-based synapses \citep{Liley2002, Suffczynski2004},  the overwhelming majority of mesoscopic neural models of brain dynamics use current-based synapses, as these are more mathematically tractable. However, it has been shown through numerical simulations \cite{Pinotsis2013a} and using spiking models \cite{Meffin2004, Richardson2004, Cavallari2014, Kuhn2004} that more biophysically realistic conductance-based synaptic mechanisms have a significant effect on the neural dynamics, which is not captured by current-based approximations.  
\\  \\
We perform a comparative bifurcation analysis to investigate the relationship between synaptic coupling and neural dynamics at the population level.  A mathematical technique called the homotopy continuation method \citep{Alexander1978}, enables us to construct a mesoscopic neural model that encapsulates both current-based and conductance-based synapses. This is performed by introducing a homotopy parameter, $h$, that continuously deforms the synaptic model between a current-based and conductance-based one.  This analysis elucidates some of the short-comings of the current-based synapse neural model.  Specifically, the analysis shows that the bifurcation structures of the parameter space in the current and conductance-based models are qualitatively different. We examine the effects of each synaptic coupling mechanism on the neural dynamics and propose an alternative highly plausible biophysical mechanism of seizure transitions unique to conductance-based synapses.  
\\
Mesoscopic models of epilepsy typically use bifurcations to explain changes in brain states  \cite{Kaczmarek1977, LopesdaSilva2003, Breakspear2006c, Robinson2002}, such as the transition to seizure found in EEG recordings.  The most common type of bifurcation used to describe the transition to seizure is a Hopf bifurcation \cite{Stefanescu2012, Breakspear2006c, Grimbert2006, Rodrigues2009, Wendling2002}, which describes a mathematical transition from a fixed point to oscillatory activity. The bifurcation parameters are typically the external input $\phi_x$ that drives the system and the network balance $\Psi$ that describes the ratio of excitatory and inhibitory activity.  The separatrices define where the model transitions are mapped in this parameter space via a bifurcation analysis.  In this study, we show that the choice of synapse in a mesoscopic neural model fundamentally changes the bifurcation structure and, consequently, the biophysical mechanism that describes epileptic transitions.  The homotopic mapping used provides a insightful means to examine these differences.  
\\ 
\textit{Model and methods:}
The mesoscopic neural model \cite{Robinson1997, jirsa1997derivation} considered here has been reformulated and modified in order to accommodate multiple synaptic coupling mechanisms via a homotopic mapping. It is specified by an input, Eq.(\ref{eq:inputeq}), an output, Eq.(\ref{eq:outputeq}), and a nonlinear coupling function relating these two quantities, Eq.(\ref{eq:sigmoid}): 
\begin{align}
\frac{dV(t)}{dt} &= \frac{-V(t)}{\tau_{1}} + \frac{{I^{\mathrm{syn}}(t) }}{C}\label{eq:inputeq}, 
\\
\left[\frac{1}{\gamma}+\frac{d}{dt}\right]^2 \phi(t)&= Q[V(t)] \label{eq:outputeq} ,
\\
Q[V(t)] &= \frac{Q_{\mathrm{max}}}{1+\exp\{-[V(t)-\theta]/\sigma\}}, \label{eq:sigmoid}
\end{align}
where $V(t)$ is the average neural membrane potential of a population of neurons, whose scale is chosen so that the zero corresponds to the reversal potential of the leak current, $\tau_1$ and $C$ are the time-constant and capacitance of the neural membrane respectively. ${I^{\mathrm{syn}}(t)}$, is the average total synaptic current, which describes the form of the synaptic input: either current-based $I^{\mathrm{cur}}(t)$ or conductance-based $I^{\mathrm{con}}(t)$. This is a sum of synaptic currents $I^{\mathrm{syn}}(t)=\sum{N_b I_b^{\mathrm{syn}}(t)}$ where $N_b$ is the number of incoming connections from population $b=e,i,x$, recurrent excitatory and inhibitory, and external neural populations, respectively. Eq.(\ref{eq:outputeq}) is derived from canonical neural field equations \cite{jirsa1997derivation, Robinson1997} and describes the propagation of a spatially uniform scalar field of firing rates $\phi(t)$ with dampening rate $\gamma$, where $Q[V(t)]$ is a sigmoidal coupling function defined in Eq.(\ref{eq:sigmoid}), which couples the input and output equations. Here, $Q_{\mathrm{max}}$ is the maximum firing rate and $\theta$ and $\sigma$ are the midpoint and spread of the sigmoidal function, respectively \cite{Freeman1972}. 
\\
The model assumes that the cortical area is on a millimeter mesoscopic scale and is spatially homogeneous and isotropic.  This is typically assumed in the literature for mesoscopic neural models of epilepsy \cite{Breakspear2006c, Rodrigues2009, Robinson2002},  and consequently removes the spatial dependence and the Laplacian term usually found in Eq.(\ref{eq:outputeq}).  The local connectivity approximation is also assumed \citep{Robinson2002, Breakspear2006c} where excitatory and inhibitory populations have the same characteristics \citep{Meffin2004}; i.e., $Q(V_e)=Q(V_i)\Rightarrow V_e=V_i=V \Rightarrow \phi_e=\phi_i=\phi$, since treating them separately produces similar results and doubles the number of free parameters \citep{Brunel2000, Meffin2004}.  
\\
\textit{Conductance-based synapses,} $I^{\mathrm{syn}}(t)$=$I^{\mathrm{con}}(t)$: This is a biophysically derived synaptic mechanism  modelled according to Ohm's law \citep{Hodgkin1952}.  The synaptic input $\phi_b(t)$ drives a transient conductance $g_b(t)$ Eq.(\ref{eq:conduc_syn1}), which is multiplicatively modulated by a membrane potential $V(t)$ term, making Eq.(\ref{eq:conduc_syn}) bilinear in $g_b(t)$ and $V(t)$. Hence, compared to current-based synapses, these are \emph{nonlinear and multiplicative}: 
\begin{align}
{I}_{b}^{\mathrm{con}}(t)&= -g_b(t)[V(t)-E_b], \label{eq:conduc_syn} \\
\dot{g}_b(t)&=-g_b(t)/\tau_2+G_b \phi_b(t), \label{eq:conduc_syn1}
\end{align}
where $G_b$ is the maximal conductance amplitude and $E_b$ is the corresponding reversal potential, which is determined by the electrostatic charge and concentrations of ionic species particular to a synapse type.  
\\
\textit{Current-based synapses,} $I^{\mathrm{syn}}(t)$=$ I^{\mathrm{cur}}(t)$: Instead of modelling the transient conductance, almost all mesoscopic neural models approximate this as a current injection, neglecting the membrane potential dependence. The generation of a post-synaptic potential from an incoming synaptic input from population $b$ is \emph{linear and additive}: 
\begin{align}
\dot{I}_{b}^{\mathrm{cur}}(t)= -I_b^{\mathrm{cur}}(t)/\tau_2 + A_b \phi_b(t), \label{eq:current_syn}
\end{align}
where $\tau_2$ is the synaptic time-constant and $A_b$ is the maximal current amplitude. 
\\
If the time-dependent membrane potential $V(t)$ is replaced by its time-independent mean, $\bar{V}$, $I_b^{\mathrm{con}}(t)$=$-g_b(t)[\bar{V}-E_b]$, then Eq.(\ref{eq:conduc_syn}) becomes the same as Eq.(\ref{eq:current_syn}), where $A_b$=$-G_b (\bar{V}-E_b)$, i.e., it becomes a current-based synapses model. To calibrate the models to receive the same level of synaptic inputs, the average charge $q_b$ injected with synaptic time-constant $\tau_2$ is equated for both synaptic mechanisms \cite{Meffin2004}, $q_b=A_b \tau_2 = -\tau_2 G_b [\bar{V}-E_b]$. This is used to define the network balance $\Psi$ as the ratio of recurrent inhibition to recurrent excitation \cite{Meffin2004}, which can be expressed as $\Psi = N_i A_i/N_e A_e = N_i G_i(\bar{V}-E_i)/N_e G_e(\bar{V}-E_e)$. The network balance $\Psi$ and the external input $\phi_x$ are used as bifurcation parameters as typically used in mesoscopic neural models of epilepsy, since changes in them are associated with pathological neurological conditions \cite{Breakspear2006c}.  Depending on the network balance, increasing the external drive excites the system so that it can transition into a seizure state \cite{Breakspear2006c, Robinson2002, Spiegler2010}. 
\\
\textit{Homotopic mapping between synaptic mechanisms:}  \label{sec:homomap}
\noindent We extend the neural model introduced to encapsulate both synaptic mechanisms by defining the synaptic current term to contain a homotopy parameter $h$, such that when $h$=0, the model has current-based synapses, and when $h$=1, it has conductance-based synapses. Homotopic continuation utilises a mapping between the two systems that continuously deforms one vector field into the other one. The fact that the vector fields may be continuously deformed into one another does not imply the resulting dynamics are topologically equivalent; i.e., they are homotopic but not necessarily homeomorphic. In this case, we can use a simple linear homotopic mapping to rewrite the synaptic current so that it continuously maps current-based to conductance-based synapses and vice-versa: 
\begin{align}
I_b^{\mathrm{syn}}(t,h)&= (1-h) I_b^{\mathrm{cur}}(t)+h I_b^{\mathrm{con}}(t), \label{eq:homotopic} \\
&=- g_b(t)[h(V(t)-\bar{V}) + \bar{V} - E_b], \label{eq:synaptic_current1}
\end{align}
where from Eqs.(\ref{eq:conduc_syn}, \ref{eq:current_syn}), $I_b^{\mathrm{cur}}(t)$=$-g_b(t) (\bar{V}$-$E_b)$ and $I_b^{\mathrm{con}}(t)$=$-g_b(t)(V(t)$-$E_b)$.  As can be seen from Eq.(\ref{eq:homotopic}), as the homotopy parameter $h \in [0,1]$ is continuously varied from zero to one, the function $I_b^{\mathrm{cur}}(t)$ is continuously deformed into $I_b^{\mathrm{con}}(t)$. 
The modulating term from the membrane potential $V(t)$ in Eq.(\ref{eq:synaptic_current1}) is now expressed as fluctuations around a mean $(V(t)-\bar{V})$ so that when $h{=}0$ these fluctuations are not included and $V(t) \mapsto \bar{V}$, a constant. However, when the modulating term is included with $h{=}1$, the constant mean membrane potential $\bar{V}$ terms in Eq.(\ref{eq:synaptic_current1}) cancel leaving the state variable $V(t)$. Consequently, this makes the synaptic mechanism conductance-based and bilinear.  
\\ \\
To facilitate analysis, the synaptic dynamics are assumed to be on a time-scale that is an order of magnitude smaller than the membrane dynamics \cite{Meffin2004}. This enables us to employ a time-scale separation and use the equilibrium values of the synaptic current and conductance w.r.t to the faster time-scale:  $\bar{I_b}^{\mathrm{cur}}(t)= \tau_2 A_b \phi_b(t)$ and $\bar{g_b}(t)=\tau_2 G_b \phi_b(t)$  in Eq.(\ref{eq:synaptic_current1}), while still keeping the bilinearity. This reduces the order of the system but does not change the bifurcation analysis, which is concerned with the asymptotic limit (i.e., ignoring the transient dynamics). Then Eq.(\ref{eq:inputeq}) and Eq.(\ref{eq:synaptic_current1}) can be combined to express a differential operator acting bilinearly on the membrane potential $V(t)$: 
\begin{align}
\left [ \frac{d}{dt}+\frac{1}{\tau_{h}(t)} \right ] V(t) 
&= \displaystyle\sum_{b=e,i,x} (E_b-(1-h)\bar{V}) \mu_b  \phi_b(t), \label{eq:homo_inputeq}
\\
\frac{1}{\tau_{h}(t)} &= \frac{1}{\tau_1} + h \displaystyle\sum_{b=e,i,x} \mu_b \phi_b(t),\label{eq:time_constant}
\end{align}
where $\phi_b$ are the firing rates, as defined in Eq.(\ref{eq:outputeq}) and $\mu_b=N_b \tau_2 G_b/C$ are the synaptic gains. The modulating membrane potential term $V(t)$, identified in Eq.(\ref{eq:synaptic_current1}) is now in the form of an active (i.e. input dependent), time-constant $\tau_{h}(t)$ \citep{Burkitt2001, Kuhn2004, Meffin2004}. When $h$=0, the active time-constant equals the passive time-constant, $\tau_{h=0}$=$\tau_1$, as in current-based synapses. For $h$=1,  the time-constant in Eq.(\ref{eq:time_constant}) is input-dependent and Eq.(\ref{eq:homo_inputeq}) is now bilinear as in conductance-based synapses. This is the essential difference between the two synaptic mechanisms that causes qualitatively different neural dynamics.  
\noindent \\
\textit{Bifurcation and nonlinear dynamics methods:} \label{sec:methods}
\noindent
The fixed points in terms of the state variables $(V^*, \phi^*)$ for the system of Eqs (\ref{eq:outputeq}, \ref{eq:sigmoid}, \ref{eq:homo_inputeq}, \ref{eq:time_constant}) are computed using a Newton-Raphson algorithm. The equations are then linearised around the fixed points to construct a Jacobian written in terms of the state variables $V(t)$ and $\phi(t)$. A local bifurcation analysis is performed and the eigenvalues of the Jacobian are computed. These eigenvalues determine the local stability of the full nonlinear system from the linearised system, as ensured by the Hartman-Grobmann lemma. 
\\
We perform a simultaneous bifurcation analysis and homotopic continuation between the two synaptic mechanisms with Eq.(\ref{eq:homo_inputeq}), using the homotopy parameter, $h$, as a bifurcation parameter.  As one model is continuously deformed into the other, at each value of  $h \in [0,1]$, the fixed points and their local stability is calculated. 
\\ \\
The following commonly accepted parameter values are retained throughout \cite{Robinson2004}: $\tau_1$=12ms, $\tau_2$=1.3 ms, $\theta$=13.3mV, $\sigma$=3.8mV, $Q_{\mathrm{max}}$=340s$^{-1}$, and for conductance-based parameters \cite{Meffin2004} $E_e$=$E_x$=0 mV, $E_i$=-75 mV, $\bar{V}$=-62.5 mV and $C$=0.35nF. The value for the dampening coefficient $\gamma$ is computed in the same way as \citep{Robinson2004}, using parameter values for the conduction velocity $v$ and the axonal range $r$: $\gamma$=v/r=0.3$ms^{-1}$/0.001$m$=300$s^{-1}$. To calibrate the models equally, $G_b$ is computed from values of the synaptic strength $s_b$ as used in current-based synapses \cite{Robinson2004}, $(s_e, s_i, s_x)$=(0.15, -1.3, 0.5)$\mu$Vs.  This is performed by equating charges (as above for the network balance) using $G_b$=$s_b C/[\tau_1 \tau_2 (E_b - \bar{V})]$, because $q_b=C s_b/\tau_1$. 
\begin{figure*}[!htp]
	\subfloat[Fixed point vs homotopy parameter]
	{\includegraphics[scale=0.26]{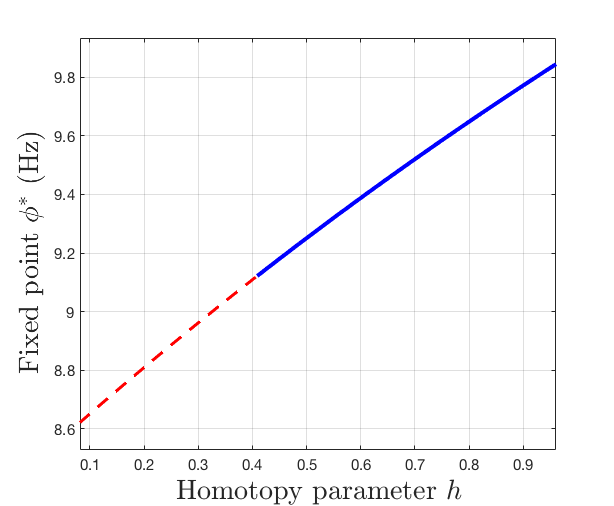}} \label{fig:homotopy_fp}
	\subfloat[Shrinking oscillatory activity]
	{\includegraphics[scale=0.255]{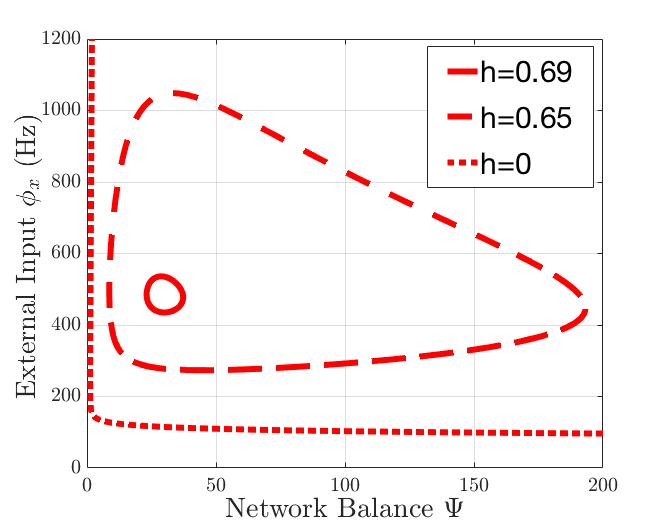}} \label{fig:shrinking}
	\subfloat[Reversal potentials bifurcation diagram]
	{\includegraphics[scale=0.255]{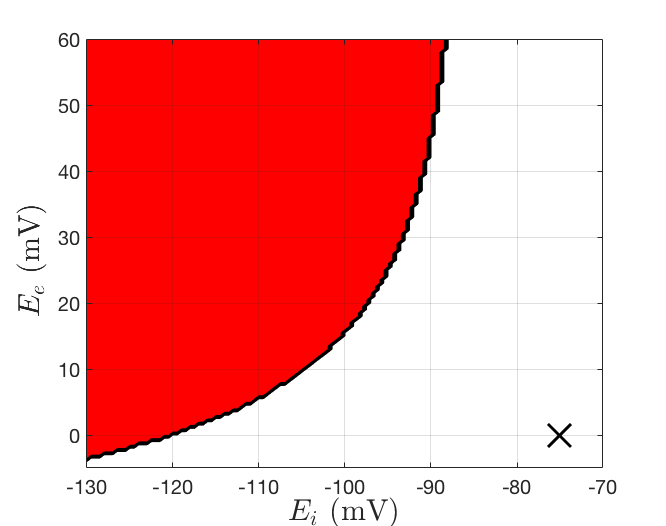}} \label{fig:reversal}
	\caption{(a) 1D bifurcation diagram showing the fixed points plotted as a function of the homotopy parameter, $h$ for a fixed value of $(\phi_x,\Psi)$=(140, 6). At a local critical value of $h_c \approx 0.408$, the system switches from a stable limit cycle (dashed) to a stable fixed point (solid). (b) 2D bifurcation diagram showing Hopf separatrices for different values of $h$ for the $(\phi_x,\Psi)$ parameter space. The red lines indicate Hopf separatrices that close, shrink and disappear at a global critical value $\hat{h}_c  \approx 0.693 \neq h_c$. (c) 2D bifurcation diagram showing Hopf separatrix in reversal potentials parameter space $(E_e, E_i)$ with $(\phi_x,\Psi)$=(140, 6). The `$\textsf{X}$' indicates the normal values used in this paper.} \label{fig}
	\setlength{\belowcaptionskip}{100pt}
\end{figure*}
\noindent \\ \\
\textit{Bifurcation analysis:} \label{sec:results} A region of parameter space is identified where the current-based model oscillates by tuning the external input $\phi_x$ and the network balance $\Psi$, as typically performed in neural models of epilepsy to generate a seizure state \cite{Breakspear2006c, Wendling2002, Suffczynski2004}. The results of a bifurcation analysis plotting the homotopy parameter $h$ vs the fixed point $\phi^*$ are shown in Fig.(\ref{fig}a).  It is found that the oscillations are suppressed for a local critical value of the homotopic parameter $h$=$h_c$$\approx$0.408 for a particular point in the same parameter subspace $(\phi_x, \Psi)$.  At the critical point, there is a bifurcation with the fixed point becoming unstable and a stable limit cycle appearing; i.e., a transition from seizure-like behaviour to normal or resting state behaviour. This is due to feedback modulation from the membrane potential term $V(t)$ in conductance-based synapses suppressing the oscillatory activity produced by the current-based synapses. Equivalently, when the increased external input to the system causes the input dependent time-constant to decrease, consequently inhibiting the system from transitioning. Hence, for the neural model with conductance-based synapses, a Hopf bifurcation is not generated for the same parameter space.  
\\  \\
In Fig.(\ref{fig}b) the oscillatory behaviour is shown as Hopf separatrices in 2D ($\phi_x$, $\Psi$) parameter space as the system is continuously deformed from current-based to conductance-based synapses. At $h$=0, current-based synapses generate an open hyperbolic Hopf separatrix curve, which as it is continuously mapped to conductance-based synapses, topologically closes and then shrinks to a mathematical point as the homotopy parameter reaches a global critical value $h=\hat{h}_c \neq h_c$.  This global critical value, $\hat{h}_c \approx$ 0.693, is for the entire $(\phi_x, \Psi)$-parameter subspace shown in Fig.(\ref{fig}b), as opposed to $h_c$, which is for a single point in the $(\phi_x, \Psi)$-subspace, as shown in Fig.(\ref{fig}a). Again, this mechanism results from multiplicative modulation of the membrane potential expressed as a contraction of the active time-constant that suppresses the oscillatory activity. This represents explicit mathematical proof that conductance-based synapses exhibit qualitatively different neural dynamics to current-based synapses, due to the absence of any transition to oscillatory dynamics, i.e., a seizure-like state.  
\noindent \\
Note that the values of the critical homotopy points, both local and global $h_c, \hat{h}_c$, are also dependent upon other parameter values such as the reversal potentials $E_e, E_i$, external input $\phi_x$, and network balance $\Psi$.  Hence, these figures actually represent a 2-D cross-section of a 5D parameter space, where if the reversal potentials are also varied it is possible for the conductance-based model to undergo a Hopf bifurcation. A bifurcation or seizure transition happens in the conductance-based model for plausible but abnormal values of the reversal potentials, Fig.(\ref{fig}c), and has been observed experimentally \citep{Ziburkus2012}. These changes are determined by ion concentration dynamics, which in turn are governed by many complex non-synaptic processes.  Biophysically, an abnormal change in the reversal potentials is a fundamentally different and highly plausible mechanism for seizure transition that cannot be described by current-based synapses.  Fig.(\ref{fig}c) shows the case where the reversal potentials $E_e, E_i$ are depolarised and hyperpolarised, respectively, during seizures \cite{Khirug2010, Holmgren2010}, $E_i$ has also been shown to become depolarised \cite{Lillis2012, Huberfeld2007}. 
\noindent \\
\textit{Discussion and conclusion:} \label{sec:conc} 
\noindent
These results provide an explicit examination of the differences between current-based and conductance-based synaptic coupling mechanisms in a mesoscopic neural model. This research extends the analytical work performed in spiking models \cite{Cavallari2014, Meffin2004, Richardson2004, Kuhn2004} and the numerical simulations in neural mass and field models \cite{Pinotsis2013a} to perform a comparative bifurcation analysis of both synaptic mechanisms. Specifically, changes in the local stability were examined by reformulating conductance-based synapses as an active time-constant and using the homotopy parameter as a bifurcation parameter.  This explicitly shows how the membrane potential modulating term, expressed as an active time-constant, has a qualitatively different effect upon the neural dynamics in contrast with its' current-based counterpart. In the current-based model, increasing the drive of the external input parameter $\phi_x(t)$ generates a Hopf bifurcation, which can be interpreted as a transition to a seizure-like state \citep{Breakspear2006c, Robinson2002, Wendling2002, Rodrigues2009}. In comparison, regardless of the network balance value, increasing the external input in the conductance-based model has no such effect, due to having a completely different bifurcation structure.  Further, we also showed that the conductance-based model transitions into an equivalent seizure-like state for abnormal values of the reversal potentials.  As discussed, changes in the reversal potentials are determined by fundamentally different, highly plausible biophysical mechanisms that cannot be captured by current-based synapses.  
\\ \\
The fluctuations of the membrane potential in a population of neurons are largely proportional to the synaptic background activity \citep{Shadlen1994, Destexhe2001}. This is reflected as an increased leakiness of the membrane that leads to a reduction of the active time-constant \citep{Burkitt2001, Kuhn2004}. When there is a fluctuating noisy input that is both inhibitory and excitatory, then the reduced time-constant is indicative of a `high conductance state' \citep{Destexhe2001}. In this state, the variability of neuronal firing and response to background input, as well as dendritic integration, is increased. Hence, although the homotopy parameter $h$ is mathematical and does not have a literal physical interpretation, it can be understood physiologically as a response to changes in the synaptic background activity that affects the conductance state of the network.  Other anatomical and physiological factors that determine when the synaptic current can be approximated as an injected current, include the spatial position \cite{TimVogels} and size of the synapses \cite{RothRossum2009} relative to the soma, and receptor type \cite{CookJohnston1999}.  Importantly, the inhibitory shunting effect of conductance-based synapses has a significant affect on the network dynamics \cite{VogelsAbbott2005} that cannot be approximated by current-based synapses \cite{Cavallari2014, Meffin2004, Richardson2004, Kuhn2004}. It is the multiplicative effect of the synaptic background fluctuations that suppresses the transition to seizure-like activity and need to be taken into account in mesoscopic neural models in order that they provide an accurate account of these phenomena. If these fluctuations are of reasonably small amplitude, for example during resting state behaviour, then current-based synapses can be an adequate approximation. However, if these fluctuations are larger in amplitude and more frequent, as typically found in electrophysiological phenomena such as oscillatory and seizure-like activity, then conductance-based synapses need to be included to provide a qualitatively accurate description of the system dynamics \citep{Richardson2007, Richardson2010}. 
\\ \\
In summary, we have constructed a homotopic mapping between two different synaptic coupling mechanisms and examined their effects on the neural dynamics of a typical mesoscopic model. The crucial finding is that the bifurcation structure of the parameter space for the different synaptic mechanisms is qualitatively different.  Further, we have suggested an alternative highly plausible biophysical mechanism for seizure transition that cannot be modelled with current-based synapses.  These results call into question the validity of previous results generated by neural models that model brain state transitions as a bifurcation and use current-based synapses. This is particularly so for models of epilepsy, as a more accurate biophysical account generates fundamentally different results.  
\\ \\
%
The authors thank Larry Abbott, John Rinzel, John Terry and Alan Lai for their insightful comments. 
We acknowledge funding support from the Australian Research Council (ARC) (LP0560684) and St.Vincent's Hospital, Melbourne.  ANB and HM acknowledge support from the ARC (DP140102947, CE140100007 resp.).  
%
\bibliography{thesis}
\end{document}